\documentclass[onecollarge]{svjour2}       % onecolumn "king-size"
\smartqed  % flush right qed marks, e.g. at end of proof
\usepackage{psfrag,graphicx,epsfig,color}

\journalname{Journal of Statistical Physics}
\begin{document}

\title{document}
\title{A model for the shapes of advected triangles}

\author{Alain Pumir and Michael Wilkinson}

\institute{Alain Pumir \at  
             Laboratoire de Physique,\\
             Ecole Normale Sup\' erieure de Lyon,\\
             F-69007, Lyon,\\
             France\\
             \email{alain.pumir@ens-lyon.fr}
            \\
             Michael Wilkinson \at
              Department of Mathematics and Statistics, \\
              The Open University,
              Walton Hall, \\
              Milton Keynes, MK7 6AA, \\
              England.\\
              \email{m.wilkinson@open.ac.uk}}

\date{Received: date / Accepted: date}

\maketitle

\begin{abstract}

Three particles floating on a fluid surface define a triangle. 
The aim of this paper is to characterise the shape
of the triangle, defined by two of its angles, as the three vertices are 
subject to a complex or turbulent motion. 
We consider a simple class of models 
for this process, involving a combination of a random strain of the fluid
and Brownian motion of the particles. Following D. G. Kendall, we map the 
space of triangles to a sphere, whose equator corresponds to degenerate 
triangles with colinear vertices, with equilaterals at the poles. 
We map our model to a diffusion process on the surface of the sphere and find 
an exact solution for the shape distribution. 
Whereas the action of the random strain tends to make the shape of 
the triangles infinitely elongated, in the presence of 
a Brownian diffusion of the vertices, the model has an equilibrium distribution of shapes. 
We determine here exactly this shape distribution in the simple case where 
the increments of the strain are diffusive. 


% 05.40.-a Fluctuation phenomena, random processes, noise, and Brownian motion
% 05.45.-a Nonlinear dynamics and nonlinear dynamical systems

\end{abstract}

\section{Introduction}
\label{sec: 0}

Consider motion of three tracer particles floating on the surface 
of a fluid which exhibits a complex flow \cite{Cas+01,Chau+11}. The three points form a 
triangle whose shape is specified by two angles, and
is therefore specified by a point in a two-dimensional manifold. 
It is of interest to consider the evolution of 
the shape of this 
triangle, independent of its scale, rotation and translation of its origin,
because this is a means to characterise the essential geometry of the flow. 
Early investigations of random shapes were mathematical  
\cite{Ken77,Ken84,Ken98}, and originally motivated by 
discerning whether nearly co-linear arrangements of archelogical 
features occured by chance or by design \cite{Ken77}.  
In the context of fluid turbulence,  models for the evolution of shapes of 
triangles
and tetrahedra have been extensively studied in the past decade in 
the fluid literature: see 
\cite{Cas+01,Mydl+98,Shr+98,PSC+00,Shr+00,FGV+01}. 
Our objective here is to define and analyse the 
motion of triplets of points in a simple model of turbulent flows.
We obtain an exact expression for the steady-state 
statistics of the triangle shapes. 

Our model flow is motivated by considering a scale decomposition of turbulent
 velocity fields. The flow at scales much larger than 
the size of the triangle merely translates and rotates the three points of
the triangle, without any shape modification. Structures existing at 
a scale comparable to triangle size are modelled by applying a
random matrix of unit determinant, to respect the incompressibility 
constraint. We refer to this motion as random strain (although
the motion involves both strain and vorticity). Lastly, 
there may also be
small-scale components in the flow which result in apparently 
uncorrelated motion of  particles.
We solve a model 
incorpoarating these effects, which is
a simplification of a model introduced in \cite{Cas+01,PSC+00}. 
Specifically, it is assumed that the particles evolve under the combined 
influence of a random strain (delta-correlated in time) and of 
independent Brownian motions.  
Because the displacement caused by a uniform strain is proportional 
to the scale size of the triangle, the strain will become dominant in the 
long-time limit as the separation of the particles grows. 
We therefore 
analyse a model where the Brownian displacement scales with the radius 
of gyration of the triangle. That is, we analyse a model for the 
small displacements $\delta \mbox{\boldmath$x$}_i$ of three particles
with positions $\mbox{\boldmath$x$}_i$ which has the following structure:
\begin{equation}
\label{eq: 0.1}
\delta \mbox{\boldmath$x$}_i = \delta {\bf A}\mbox{\boldmath$x$}_i
+R \delta \mbox{\boldmath$r$}_i
\end{equation}
where $\delta {\bf A}$ is a traceless random symmetric matrix with diffusive fluctuations,
$\delta \mbox{\boldmath$r$}_i$ is a Brownian displacement, and $R$ is the radius of gyration, defined by
\begin{equation}
\label{eq: 0.2}
R^2=\frac{1}{6}\sum_{i=1}^3\sum_{j=1}^3 |\mbox{\boldmath$x$}_i-\mbox{\boldmath$x$}_j|^2
\ .
\end{equation}
The stochastic incrememts $\delta {\bf A}$ and $\delta \mbox{\boldmath$x$}_i$ in equation
 (\ref{eq: 0.1}) are both multiplied by terms which depend upon the coordinates 
$\mbox{\boldmath$x$}_i$. This implies a potential source of ambiguity about whether the 
increment $\delta \mbox{\boldmath$x$}_i$ at each time step should be evaluated using the It\=o 
or the Stratonovich rule \cite{vKa81}. We will show that this ambiguity only affects the way the
size of the triangle grows in time. When considering the evolution of the shape of
the triangle, we do not need to distinguish whether the stochastic increments are of It\=o or
Stratonovich type.

In this article, we begin by investigating separately the action 
of the strain and of the 
Brownian terms. We then consider the two terms together, and
find an exact expression for the distribution of the shapes
of the triangles which are generated by this random process.

Specifying the shape of a triangle requires two parameters, which could be
two of the three angles. For our purposes, however, another parametrisation
is much more useful. D.G. Kendall argued that it is natural to describe the 
shape of a triangle by a point on the surface of a sphere \cite{Ken77}, 
which we will refer to as the Kendall sphere. His paper also states that if the 
vertices of a triangle execute independent Brownian motions, then the image 
of this process is uniform diffusion on the surface of the Kendall sphere. 
His original paper \cite{Ken77} does not give a derivation of this result, and derivations which we have
seen in later works rely upon computer algebra packages \cite{Ken98} or upon quite
abstract machinery \cite{Ken84}. In section \ref{sec: 1} we start by explaining the Kendall
sphere,  its relation to other parametrisations, and we describe a Langevin equation
which will prove to be a convenient representation of diffusion on the surface
of a sphere. In section \ref{sec: 3} we define the ensemble of random strain $\delta {\bf A}$
introduced in (\ref{eq: 0.1}), and evaluate the image of this random strain process
on the surface of the Kendall sphere. 
We show that this leads to a diffusion process 
which is axially symmetric on the sphere. In section \ref{sec: 2} we use the same 
framework to give a direct demonstration of Kendall's result
that the image of Brownian motion of the three points in Cartesian space is 
unbiased, isotropic and homogeneous diffusion on the Kendall sphere.

 In section \ref{sec: 4} we combine
the results of sections \ref{sec: 3} and \ref{sec: 2} to determine a Fokker-Planck
equation which determines the probability density for the model 
(\ref{eq: 0.1}) on the Kendall sphere. We solve
this equation for the steady-state shape distribution.

The analysis in sections \ref{sec: 3}-\ref{sec: 4} is uses the strategy of determining stochastic
increments of polar coordinates on the Kendal sphere, and using these to surmise the Fokker-Planck
equation. A complementary approach is to develop a Fokker-Planck equation for the model (\ref{eq: 0.1})
directly, and make a change of variable. This approach is explored in section \ref{sec: 5}, where we also 
show how the Fokker-Planck operator can be expressed as a linear combination of Casimir operators. 
Last, we briefly summarize our results in Section \ref{sec: 6}.

\section{The Kendall sphere}
\label{sec: 1}

Consider a triangle in the plane formed by three points 
$\mbox{\boldmath$x$}_i$, $i=1,2,3$.
The shape of the triangle is described by two two-component vectors $\mbox{\boldmath$u$}_1$ and
$\mbox{\boldmath$u$}_2$ and by a $2\times 2$ matrix ${\bf w}$, defined by
\begin{equation}
\label{eq: 1.1}
\mbox{\boldmath$u$}_1=\frac{1}{\sqrt{2}}\left(\mbox{\boldmath$x$}_1-\mbox{\boldmath$x$}_2\right)
\ ,\ \ \ 
\mbox{\boldmath$u$}_2=\frac{1}{\sqrt{6}}(2\mbox{\boldmath$x$}_3-\mbox{\boldmath$x$}_1-\mbox{\boldmath$x$}_2)
\end{equation}
and 
\begin{equation}
\label{eq: 1.2}
{\bf w}=
\left(\begin{array}{cc}
u_{1,1} & u_{2,1} \cr
u_{1,2 }& u_{2,2}
\end{array}\right)
\end{equation}
where $u_{i,a}$, $a=1,2$ are the Cartesian components of $\mbox{\boldmath$u$}_i$. 
It is natural to represent the matrix ${\bf w}$ by a singular-value decomposition, writing
\begin{equation}
\label{eq: 1.3}
{\bf w}={\bf R}(\chi)\,{\rm diag}(\lambda_1,\lambda_2)\,{\bf R}(\phi/2)
\end{equation}
where ${\bf R}(\theta)$ is the matrix describing a rotation by an angle $\theta$:
\begin{equation}
\label{eq: 1.4}
{\bf R}(\theta)=\left(\begin{array}{cc}
\cos\theta & -\sin\theta \cr
\sin\theta & \cos\theta
\end{array}\right)
\ .
\end{equation}
Note that the angle $\chi $ in (\ref{eq: 1.3}) describes an overal rotation of the triangle,
and is irrelevant to describing its shape. 
The radius of gyration $R$ of the triangle may be expressed in terms of the singular values $\lambda_1$ and $\lambda_2$:
\begin{equation}
\label{eq: 1.5}
R^2=\sum_{i=1}^3|\mbox{\boldmath$x$}_i-\bar{\mbox{\boldmath$x$}}|^2=\mbox{\boldmath$u$}_1^2+\mbox{\boldmath$u$}_2^2
={\rm tr}({\bf w}^{\rm T}{\bf w})=\lambda_1^2+\lambda_2^2
\end{equation}
where $\bar{\mbox{\boldmath$x$}}$ is the centre of mass of the three points. 
The area of the triangle is simply expressed in terms of $\lambda_1$ and $\lambda_2$ as: 
\begin{equation}
\label{eq: 1.55}
\zeta={\rm det}({\bf w})=\lambda_1\lambda_2
\end{equation}
We note that the size of the triangle can be characterized either by 
$R$ or by $\zeta$.
Using $\zeta$ is natural when considering area-preserving dynamics
\cite{Shr+98}. 
However, because the parametrization involving $\zeta$ has a 
singularity when the three points defining the triangle become co-linear,
it is more natural to use $R$ instead of $\zeta$ when the dynamics allows the vertices to
become co-linear. For this reason we prefer $R$ over $\zeta$ as the scale coordinate throughout
most of this paper. 
The angle $\phi$ characterizes the rotation in the space of triangles with 
a given set of values of $\lambda_1$ and $\lambda_2$. We note that a cyclic 
permutation
of the vertices  $\mbox{\boldmath$x$}_i$, $(1,2,3) \rightarrow (2,3,1)$
(respectively $(1,2,3) \rightarrow (3,1,2)$) amounts to a transformation
$\phi \rightarrow \phi + 2 \pi/3$ (respectively $\phi \rightarrow \phi + 4 \pi/3$). 

The shape of the 
triangle may be specified by the variables $\phi$ and
\begin{equation}
\label{eq: 1.6}
\Lambda=\frac{\lambda_1}{\lambda_2}
\end{equation}
instead of by two angles. 

It is useful to consider the form of the matrix ${\bf w}^{\rm T}{\bf w}$:
\begin{eqnarray}
\label{eq: 1.7}
{\bf w}^{\rm T}{\bf w}&=&
\left(\begin{array}{cc}
\mbox{\boldmath$u$}_1^2 & \mbox{\boldmath$u$}_1\cdot \mbox{\boldmath$u$}_2 \cr
\mbox{\boldmath$u$}_1\cdot \mbox{\boldmath$u$}_2 & \mbox{\boldmath$u$}_2^2
\end{array}\right)
\nonumber \\
&=&{\bf R}(-\phi/2) {\rm diag}(\lambda_1,\lambda_2){\bf R}(-\chi)
{\bf R}(\chi){\rm diag}(\lambda_1,\lambda_2){\bf R}(\phi/2))
\nonumber \\
&=&{\bf R}(-\phi/2){\rm diag}(\lambda_1^2,\lambda_2^2){\bf R}(\phi/2)
\nonumber \\
&=&\lambda_2^2
\left(\begin{array}{cc}
\frac{\Lambda^2+1}{2}+\frac{\Lambda^2-1}{2}\cos\phi & -\frac{\Lambda^2-1}{2}\sin\phi \cr
-\frac{\Lambda^2-1}{2}\sin\phi & \frac{\Lambda^2+1}{2}-\frac{\Lambda^2-1}{2}\cos\phi
\end{array}\right)
\end{eqnarray}
so that the geometrical properties of the vectors $\mbox{\boldmath$u$}_1$ and $\mbox{\boldmath$u$}_2$ have period $2\pi$ in the angle $\phi$.

It is convenient to transform the ratio $\Lambda$ into another variable, $z$,
defined by 
\begin{equation}
\label{eq: 1.8}
z=\frac{2{\rm det}({\bf w})}{{\rm tr}({\bf w}^{\rm T}{\bf w})}=\frac{2\Lambda}{1+\Lambda^2}
=\frac{2\zeta}{R^2}
\ .
\end{equation}
Note that the variable $z$, denoted $w$ in \cite{Cas+01,Shr+98},
satisfies  $-1\le z\le +1$, so that a point the region $|z|\le 1$, $0\le \phi<2\pi$ uniquely defines
the shape of a triangle. The points $z=\pm 1$ correspond to equilateral triangles, with their
shape being independent of the angular coordinate $\phi$. We can therefore regard the space of triangles
as being equivalent to the surface of a sphere with azimuthal angle $\phi$, and with polar angle 
$\theta$, related to $z$ by $z=\cos\theta$. The equator corresponds to degenerate triangles consisting 
of three colinear points. The poles are equilateral triangles. There are three singular points on the equator,
at which two points of the co-linear denerate triangle coincide. This device of representing the 
shape of a triangle by a point on the surface of a sphere was first introduced 
by D. G. Kendall \cite{Ken77}. We will refer to it as the Kendall sphere.

In the case where a triangle is subjected to a sequence of uncorrelated 
infinitesimal perturbations, the representative point diffuses over the surface
of the sphere. We will characterise this diffusion process by specifying 
the stochastic increments $\delta z$ and $\delta \phi$ of the coordinates $z$ and $\phi$. 
In general the diffusion process may be inhomogeneous, 
it may have an anisotropic diffusion coefficient, and it may have drift
terms. As a reference model, let us consider the case of homogeneous
and isotropic diffusion on the Kendall sphere, with diffusion coefficient $D$.
We characterise this process in terms of the stochastic increments of $z$
and $\phi$.

To characterise this homogeneous diffusive evolution let us use a system of standard Cartesian
coordinates: $(x,y,z)=(\sin\theta\cos\phi,\sin\theta\sin\phi,\cos\theta)$.
Without loss of generality we can use invariance with respect to 
the azimuthal angle $\phi$, and consider a point where $y=0$. Consider an orthogonal basis
defined at the point $(\sin\theta,0,\cos\theta)$:
\begin{eqnarray}
\label{eq: 1.9}
{\bf n}&=&(\sin\theta,0,\cos\theta)=(x,0,z)
\nonumber \\
{\bf m}&=&(-\cos\theta,0,\sin\theta)=(-z,0,x)
\nonumber \\
{\bf k}&=&(0,1,0)
\ .
\end{eqnarray}
The diffusive fluctuation of a point ${\bf n}=(x,0,z)$ on the surface is
\begin{equation}
\label{eq: 1.10}
\delta {\bf n}=\delta X{\bf m}+\delta Y{\bf k}+\delta Z{\bf n}
\end{equation}
where $\delta X$ and $\delta Y$ are angular increments, satisfying $\langle \delta X\rangle=\langle \delta Y\rangle=0$ and
\begin{equation}
\label{eq: 1.11}
\langle \delta X^2\rangle=\langle \delta Y^2\rangle=2D\delta t
\ ,\ \ \ \ 
\langle \delta X\delta Y\rangle=0
\ .
\end{equation}
The value of $\delta Z$ is determined from $\delta X$ and $\delta Y$ 
using the constraint
that ${\bf n}$ remains normalised: this implies that $\delta Z=-\frac{1}{2}(\delta X^2+\delta Y^2)$. 
Consider the increments of $z$ and $\phi$ under this process: we have
$\delta Y=\sin\theta \delta \phi$ and $({\bf n}+\delta {\bf n}).{\bf e}_3=z+\delta z
=\cos\theta-\sin \theta \delta X+\cos\theta \delta Z$.
Solving for $\delta \phi$, $\delta z$, we find
\begin{eqnarray}
\label{eq: 1.14}
\delta \phi&=& \frac{1}{x}\delta Y
\nonumber \\
\delta z&=&-x\delta X-\frac{1}{2}z(\delta X^2+\delta Y^2)
\end{eqnarray}
where $z=\cos\theta$, $x=\sqrt{1-z^2}$.  The diffusion and drift coefficients for homogeneous 
diffusion on the sphere are defined by writing $\langle \delta z\rangle=v_z\delta t$, $\langle \delta z^2 \rangle=2D_z\delta t$, 
$\langle \delta \phi^2\rangle=2D_\phi \delta t$, are therefore
\begin{eqnarray}
\label{eq: 1.15}
D_z&=&(1-z^2)D
\nonumber \\
D_\phi&=&\frac{D}{1-z^2}
\nonumber \\
v_z&=&-2zD
\nonumber \\
v_\chi&=&0
\ .
\end{eqnarray}
This representation of homogeneous, isotropic diffusion on the surface of a sphere 
will prove to be an instructive comparison case for the random strain model
which we consider next. 

\section{Image of random strain on the Kendall sphere}
\label{sec: 3}

Here we consider the dynamics of the three verices $\mbox{\boldmath$r$}_i$ of the triangle under 
the action of a matrix ${\bf M}$ which satisfies ${\rm det}({\bf M})=1$.
We shall describe the shape of a triangle by the matrix used to obtain it by deformation
of an equilateral triangle. Call this matrix ${\bf M}$. We consider area-preserving 
flows, so that we require ${\rm det}({\bf M})=1$. The equilateral triangle (with unit radius of gyration) is described
by
\begin{equation}
\label{eq: 3.1}
{\bf w}_0=\frac{1}{\sqrt{2}}\left(\begin{array}{cc}
1 & 0 \cr 
0 & 1
\end{array}\right)
\ .
\end{equation}
A general triangle is described by 
\begin{equation}
\label{eq: 3.2}
{\bf w}={\bf M}\,{\bf w}_0
\ .
\end{equation}
We use a singular-value decomposition representation of ${\bf M}$:
\begin{equation}
\label{eq: 3.3}
{\bf M}={\bf R}(\chi)\ {\rm diag}(\lambda_1,\lambda_2)\ {\bf R}(\phi/2)
\end{equation}
with $\lambda_1\lambda_2=1$.

The strain term, introduced in (\ref{eq: 0.1}), acts multiplicatively on 
the variables defining the shape ${\bf M}$, in the form:
\begin{equation}
\label{eq: 3.4}
{\bf M}'=({\bf I}+\delta {\bf A})\,{\bf M}={\bf R}(\chi+\delta \chi)\,{\rm diag}(\lambda_1+\delta \lambda_1,\lambda_2+\delta \lambda_2)\,{\bf R}(\phi/2+\delta \phi/2)
\end{equation}
where $\delta {\bf A}$ is an infinitesimal random strain, of the form
\begin{equation}
\label{eq: 3.5}
\delta {\bf A}=\left(\begin{array}{cc}
\delta a & \delta b \cr
\delta b & -\delta a
\end{array}\right)
\ .
\end{equation}
The statistics of the elements of the random strain are
$\langle \delta a\rangle=\langle \delta b\rangle=0$ and 
\begin{equation}
\label{eq: 3.6}
\langle \delta a^2\rangle =2D_a\delta t
\ ,\ \ \ 
\langle \delta b^2 \rangle=2D_b\delta t
\ ,\ \ \ 
\langle \delta a\delta b\rangle =0
\ .
\end{equation}
Imposing the reqirement that the statistics of this random strain matrix are rotationally
invariant leads to the condition $D_a=D_b\equiv D_s$. An infinitesimal rotation could have been included
in the definition of the random strain (by making the off-diagonal elements dissimilar), but
this in clearly irrelevant to shape fluctuations. 

Because the noise term $\delta {\bf A}$
acts multiplicatively on the matrix ${\bf M}$ defining the shape of the 
triangle,
the precise definition of $\delta {\bf A} {\bf M}$ 
matters when practically integrating the equations. The simplest
procedure consisting in taking 
${\bf M}(t + \delta t) - {\bf M} (t)  \approx \delta {\bf A} \cdot {\bf M}(t)$ 
(the It\=o procedure), and the physically more correct procedure, consisting
in defining:
${\bf M}(t + \delta t) - {\bf M} (t)  \approx \frac{1}{2} \delta {\bf A}  \cdot ({\bf M}(t) + {\bf M}(t + \delta t) )$ 
(the Stratonovich procedure) \cite{vKa81} differ by a diagonal term,
$2 D_s \delta t ~ \bf{I}$, which dilates in a similar way the two
eigenvalues $\lambda_1$ and $\lambda_2$, thus leaving the variables
that characterize the shape, $z$ and $\phi$, unaffected. 
For this reason, the precise way of defining the model here turns out to be
immaterial.

Consider the evolution of the parameters $\phi$, $\chi$, $\lambda_1$, $\lambda_2$ under this 
random strain evolution: from (\ref{eq: 3.4}) we obtain
\begin{equation}
\label{eq: 3.7}
{\bf R}(\delta \chi)\,{\rm diag}(\lambda_1 +\delta \lambda_1,\lambda_2+\delta \lambda_2)\,{\bf R}(\delta \phi/2)
={\rm diag}(\lambda_1,\lambda_2)+\delta {\bf A}'{\rm diag}(\lambda_1,\lambda_2)
\end{equation}
where 
\begin{equation}
\label{eq: 3.8}
\delta {\bf A}'={\bf R}(-\chi)\,\delta {\bf A}\,{\bf R}(\chi)
\equiv\left(\begin{array}{cc}
\delta a' & \delta b' \cr
\delta b' & -\delta a'
\end{array}\right)
\ .
\end{equation}
In the following we use the rotational invariance of the statistics of $\delta {\bf A}$ to conclude
that the statistics of its elements satisfy 
$\langle \delta a'^2\rangle=\langle \delta b'^2\rangle=2D_s \delta t$.
Expanding the factors of equation (\ref{eq: 3.7}) up to quadratic terms in the small increments 
$\delta \chi$, $\delta \phi$ and $\delta \lambda_i$, and collecting terms, we find
\begin{eqnarray}
\label{eq: 3.10}
\lambda_1 \delta a'&=&\delta \lambda_1-\frac{\lambda_1}{8}(4\delta \chi^2+\delta \phi^2)-\frac{\lambda_2}{2} \delta \chi \delta \phi 
\nonumber \\
-\lambda_2 \delta a'&=&\delta\lambda_2-\frac{\lambda_2}{8}(4\delta \chi^2+\delta\phi^2)-\frac{\lambda_1}{2}\delta \chi \delta \phi
\nonumber \\
\lambda_2\delta b' &=&\lambda_2\delta \chi +\frac{\lambda_1}{2}\delta \phi +\frac{1}{2}\delta\lambda_1\delta \phi+\delta \lambda_2\delta \chi
\nonumber \\
\lambda_1 \delta b'&=&-\lambda_1\delta\chi -\frac{\lambda_2}{2}\delta \phi-\delta\lambda_1\delta\chi-\frac{1}{2}\delta \lambda_2\delta \phi
\ .
\end{eqnarray}
The first order relations between the increments $\delta a'$, $\delta b'$ and the 
increments of the shape parameters are therefore
\begin{eqnarray}
\label{eq: 3.11}
\delta \lambda_1&=&\lambda_1\delta a'
\nonumber \\
\delta \lambda_2&=&-\lambda_2\delta a'
\nonumber \\
\delta \phi&=&\frac{4\lambda_1 \lambda_2}{\lambda_1^2-\lambda_2^2}\delta b'
\nonumber \\
\delta \chi&=&\frac{\lambda_1^2+\lambda_2^2}{\lambda_2^2-\lambda_1^2}\delta b'
\end{eqnarray}
so that
\begin{eqnarray}
\label{eq: 3.12}
\langle \delta \chi^2\rangle&=&
\frac{2D_s\delta t (\lambda_1^2+\lambda_2^2)^2}{(\lambda_1^2-\lambda_2^2)^2}
\nonumber \\
\langle \delta \phi^2\rangle&=&
\frac{32D_s\delta t \lambda_1^2\lambda_2^2}{(\lambda_1^2-\lambda_2^2)^2}
\nonumber \\
\langle \delta \phi\delta \chi \rangle &=&
\frac{-8D_s\delta t\lambda_1\lambda_2(\lambda_1^2+\lambda_2^2)}{(\lambda_1^2-\lambda_2^2)^2}
\ .
\end{eqnarray}
Adding the corrections arising from the quadratic terms then gives
\begin{eqnarray}
\label{eq: 3.13}
\lambda_1 \delta a'&=&\delta \lambda_1-\lambda_1\frac{\lambda_1^2+3\lambda_2^2}{\lambda_1^2-\lambda_2^2}D_s\delta t
\nonumber \\
-\lambda_2\delta a'&=&\delta \lambda_2+\lambda_2\frac{\lambda_2^2+3\lambda_1^2}{\lambda_1^2-\lambda_2^2}D_s \delta t
\nonumber \\
\lambda_2\delta b' &=&\lambda_2\delta \chi +\frac{1}{2}\lambda_1\delta \phi 
\nonumber \\
\lambda_1 \delta b'&=&-\lambda_1\delta\chi -\frac{1}{2}\lambda_2\delta \phi
\ .
\end{eqnarray}
The first two of these relations are consistent with the constraint $\delta (\lambda_1\lambda_2)=0$, and the second
pair follow from $\langle \delta a'\delta b'\rangle=0$ .

Now transform to new variables $\Lambda=\lambda_1/\lambda_2$, $R=\sqrt{\lambda_1^2+\lambda_2^2}$.
It is clear that the evolution is scale invariant, so that the dynamics does not depend upon the radius 
of gyration $R$. We find
\begin{eqnarray}
\label{eq: 3.14}
\delta \Lambda&=&\frac{\delta \lambda_1}{\lambda_2}-\Lambda \frac{\delta \lambda_2}{\lambda_2}+\Lambda \frac{\delta \lambda_2^2}{\lambda_2^2}-\frac{\delta \lambda_1\delta \lambda_2}{\lambda_2^2}
\nonumber \\
&=&2\Lambda \delta a'+\frac{8\lambda_1^2 \Lambda D_s\delta t}{\lambda_1^2-\lambda_2^2}
\ .
\end{eqnarray}
The increment of $\phi$ is
\begin{equation}
\label{eq: 3.15}
\delta \phi=\frac{4\lambda_1\lambda_2}{\lambda_1^2-\lambda_2^2}\delta b'=\frac{4\Lambda}{\Lambda^2-1}\delta b'
\ .
\end{equation}
The drift and diffusion coefficients in the $\Lambda$, $\phi$ variables are:
\begin{eqnarray}
\label{eq: 3.16}
D_\Lambda(\Lambda)&=&4\Lambda^2 D_s
\nonumber \\
D_\phi(\Lambda)&=&\frac{16\Lambda^2 D_s}{(\Lambda^2-1)^2}
\nonumber \\
v_\Lambda(\Lambda)&=&\frac{8\Lambda^3 D_s}{\Lambda^2-1}
\nonumber \\
v_\phi&=&0
\ .
\end{eqnarray}
Now work in terms of $z=2\Lambda/(\Lambda^2+1)$: the diffusion and drift coefficients are 
\begin{eqnarray}
\label{eq: 3.17}
D_z(\Lambda)&=&\left(\frac{{\rm d}z}{{\rm d}\Lambda}\right)^2
D_\Lambda=\frac{4(1-\Lambda^2)^2}{(1+\Lambda^2)^4}4\Lambda^2 D_s
\nonumber \\
v_z (\Lambda)&=&\frac{{\rm d}z}{{\rm d}\Lambda}v_\Lambda+\frac{{\rm d}^2z}{{\rm d}\Lambda^2}D_s\Lambda
=\frac{-64 D_s\Lambda^3}{(\Lambda^2+1)^3}
\ .
\end{eqnarray}
To express these in terms of $z$, consider an intermediate variable $\Theta$, satisfying 
\begin{equation}
\label{eq: 3.18}
\Lambda=\frac{\sin\Theta}{\cos\Theta}
\ ,\ \ \ 
z=\sin(2\Theta)
\ .
\end{equation}
Then:
\begin{eqnarray}
\label{eq: 3.19}
D_z(z)&=&\frac{16D_s \Lambda^2(1-\Lambda^2)^2}{(1+\Lambda^2)^4}
\nonumber \\
&=&4D_s \sin^2(2\Theta)\cos^2(2\Theta)
\nonumber \\
&=&4D_s z^2(1-z^2)
\end{eqnarray}
and
\begin{eqnarray}
\label{eq: 3.20}
D_\phi(z)&=&\frac{16\Lambda^2 D_s}{\Lambda^2-1}=-\frac{4 D_s \sin^2(2\Theta)}{\cos^2(2\Theta)}
\nonumber \\
&=&\frac{4 D_s  z^2}{1-z^2}
\end{eqnarray}
finally
\begin{eqnarray}
\label{eq: 3.21}
v_z(z)&=&\frac{-32  D_s \Lambda^3}{(1+\Lambda^2)^3}
\nonumber \\
&=&-8 D_s z^3
\ .
\end{eqnarray}
It is interesting to note that the coefficients differ from those of the homogeneous 
diffusion on a sphere (considered in section \ref{sec: 1}) by a simple factor, namely $4z^2$.

\section{Image of Brownian motion on the Kendall sphere}
\label{sec: 2}

In the Introduction we mentioned that if the vertices execute independent
Brownian motion, then the dynamics of the representative point on the 
surface of the Kendall sphere is homogeneous diffusion. The published 
derivations \cite{Ken84,Ken98} are difficult to compare with the derivation 
of the diffusion process for the random strain model which we analysed 
in section \ref{sec: 3}. In this section we present an elementary derivation
of Kendall's result, using the same approach as for the random strain model.

Let the vertices of a triangle make small, independent and isotropic diffusive
displacements, as if they are undergoing Brownian motion. The vectors $\mbox{\boldmath$u$}_i$ 
undergo displacements with components
\begin{equation}
\label{eq: 2.1}
\delta u_{1,a}=\frac{1}{\sqrt{2}}(\delta x_{1,a}-\delta x_{2,a})
\ ,\ \ \ 
\delta u_{2,a}=\frac{1}{\sqrt{6}}(2\delta x_{3,a}-\delta x_{1,a}-\delta x_{2,a})
\ .
\end{equation}
The statistics of these small displacements satisfy
\begin{equation}
\label{eq: 2.2}
\langle \delta u_{i,j}\rangle=0
\ ,\ \ \ 
\langle \delta u_{i,j}\delta u_{k,l}\rangle=2\delta_{ik}\delta_{jl} D_b \delta t
\end{equation}
where $D_b $ is a diffusion coefficient. Note that $\langle \delta u_{i,j}^2\rangle =\langle \delta x_{k,l}^2\rangle$
for any choice of $i$, $j$, $k$, $l$, so that $D_b $ is the same as the diffusion coefficient for the Brownian
motion in coordinate space.

We can characterise the change in the triangle by considering a change matrix ${\bf w}$ and hence the 
changes in the variables $\phi$ and $z$. Applying rotation matrices to ${\bf w}+\delta {\bf w}$, we obtain
\begin{equation}
\label{eq: 2.3}
{\bf R}(-\chi)({\bf w}+\delta {\bf w}){\bf R}(-\phi/2)={\rm diag}(\lambda_1,\lambda_2)
+{\bf R}(-\chi)
\left(\begin{array}{cc}
\delta u_{1,1} & \delta u_{2,1} \cr
\delta u_{1,2} & \delta u_{2,2}
\end{array}\right)
{\bf R}(-\phi/2)
\equiv {\rm diag}(\lambda_1,\lambda_2)+\delta {\bf w}'
\ .
\end{equation}
The elements of $\delta {\bf w}'$ are independent of $\chi$ and $\phi$: for example
\begin{equation}
\label{eq: 2.4}
\delta w_{11}=
\cos\chi \cos(\phi/2) \delta u_{1,1}-\sin\chi\cos(\phi/2) \delta u_{1,2}
+\cos\chi\sin(\phi/2)\delta u_{2,1}-\sin\chi\sin(\phi/2) \delta u_{2,2}
\end{equation}
has a variance  $\langle (\delta w'_{11})^2\rangle=\langle (\delta w_{11})^2\rangle$
which is independent of $\phi$ and $\chi$, and the same is true of the other components
of $\delta w_{ij}$.
Because of this invariance it suffices to  perform perturbation theory about the 
$\chi=0$, $\phi=0$ case:
\begin{equation}
\label{eq: 2.6}
\left(\begin{array}{cc}
\lambda_1+\delta u'_{1,1} &  \delta u'_{2,1} \cr
\delta u'_{1,2} & \lambda_2+\delta u'_{2,2}
\end{array}\right)
=\left(\begin{array}{cc}
\cos\delta \chi & \sin\delta \chi \cr
-\sin\delta \chi & \cos\delta \chi
\end{array}\right)
\left(\begin{array}{cc}
\lambda_1+\delta \lambda_1 & 0 \cr
0 & \lambda_2+\delta \lambda_2
\end{array}\right)
\left(\begin{array}{cc}
\cos(\delta \phi/2) & \sin(\delta \phi/2) \cr
-\sin(\delta \phi/2) & \cos(\delta \phi/2)
\end{array}\right)
\end{equation}
where the $\delta u'_{i,j}$ are components of $\delta u_{i,j}$ in the rotated 
coordinate system.
Expanding in powers of $\delta \theta $ and $\delta \chi$, and retaining
terms only up to quadratic order in the small increments, we find
\begin{eqnarray}
\label{eq: 2.7}
\delta u'_{1,1}&=&\delta \lambda_1-\frac{\lambda_1}{8}(4\delta \chi^2+\delta\phi^2)-\frac{\lambda_2}{2} \delta \chi\delta \phi
\nonumber \\
\delta u'_{2,2}&=&\delta \lambda_2-\frac{\lambda_2}{8}(4\delta\chi^2+\delta\phi^2)-\frac{\lambda_1}{2}\delta\chi\delta\phi
\nonumber \\
\delta u'_{1,2}&=&-\lambda_1\delta \chi-\frac{\lambda_2}{2}\delta \phi-\frac{\delta \lambda_2\delta \phi}{2}-\delta\lambda_1\delta \chi
\nonumber \\
\delta u'_{2,1}&=&\frac{\lambda_1}{2}\delta \phi+\lambda_2\delta\chi+\frac{\delta\lambda_1\delta\phi}{2}+\delta\lambda_2\delta \chi
\ .
\end{eqnarray}
These equations can be solved to determine $\delta \lambda_i$, $\delta \theta$ and $\delta \chi$
in terms of the elements of $\delta {\bf w}'$. We follow the same approach as in section \ref{sec: 3}. 
First, compute the linear terms from the linear terms of (\ref{eq: 2.7}). We then use these 
expressions to estimate the covariances of the small increments $\delta \phi$, $\delta \chi$
and $\delta \lambda_i$. We find that the stochastic increments are
\begin{eqnarray}
\label{eq: 2.11}
\delta \lambda_1&=&\delta u'_{1,1}+\frac{2\lambda_1 D_b \delta t}{\lambda_1^2-\lambda_2^2}
\nonumber \\
\delta \lambda_2&=&\delta u'_{2,2}-\frac{2\lambda_2 D_b \delta t}{\lambda_1^2-\lambda_2^2}
\nonumber \\
\delta \phi&=&\frac{2(\lambda_1\delta u'_{2,1}+\lambda_2\delta u'_{1,2})}{\lambda_1^2-\lambda_2^2}
\nonumber \\
\delta \chi&=&\frac{\lambda_1\delta u'_{1,2}+\lambda_2\delta u'_{2,1}}{\lambda_2^2-\lambda_1^2}
\ .
\end{eqnarray}
From now on we concentrate upon the evolution of the coordinates which define the
shape: these can be taken to be $\phi$ and $\Lambda=\lambda_1/\lambda_2$. Note that
\begin{equation}
\label{eq: 2.12}
\delta \Lambda=\frac{\lambda_1+\delta \lambda_1}{\lambda_2+\delta \lambda_2}-\frac{\lambda_1}{\lambda_2}
=\frac{\delta \lambda_1}{\lambda_2}-\Lambda \frac{\delta\lambda_2}{\lambda_2}+\Lambda \frac{\delta \lambda_2^2}{\lambda_2^2}-\frac{\delta\lambda_1\delta\lambda_2}{\lambda_2^2}
\ .
\end{equation}
Now normalise the scale of the triangle so that the radius of gyration 
$\sqrt{\lambda_1^2+\lambda_2^2}=R$ is factored out of the stochastic
increments:
\begin{eqnarray}
\label{eq: 2.13}
\delta \Lambda&=&\frac{\delta u'_1}{\lambda_2}+\frac{2\Lambda D_b \delta t}{\lambda_1^2-\lambda_2^2}-\frac{\Lambda\delta v'_2}{\lambda_2}+\frac{2\Lambda D_b \delta t}{\lambda_1^2-\lambda_2^2}+\frac{2\Lambda D_b \delta t}{\lambda_2^2}
\nonumber \\
&=&\frac{1}{R}\frac{\delta u'_1}{\lambda_2 R}-\frac{\Lambda}{R}\frac{\delta v'_2}{\lambda_2 R}
+\frac{2\Lambda D_b \delta t}{R^2}\left[\frac{2R^2}{\lambda_1^2-\lambda_2^2}+\frac{R^2}{\lambda_2^2}\right]
\nonumber \\
&=&\frac{1}{R}\frac{\delta u'_1}{\lambda_2 R}-\frac{\Lambda}{R}\frac{\delta v'_2}{\lambda_2 R}
+\frac{1}{R^2}\frac{2\Lambda(\Lambda^2+1)^2 D_b  \delta t}{\Lambda^2-1}
\ .
\end{eqnarray}
We consider the evolution of the shape of the triangle with the radius of gyration frozen. 
When the radius of gyration is $R=1$, the increment of $\Lambda$ may therefore be written
\begin{equation}
\label{eq: 2.14}
\delta \Lambda=(1+\Lambda^2)\delta x+\frac{2\Lambda(1+\Lambda^2)^2 D_b \delta t}{(\Lambda^2-1)}
\end{equation}
where 
\begin{equation}
\label{eq: 2.15}
\delta x=\frac{\delta u'_{1,1}-\Lambda \delta u'_{2,2}}{\sqrt{1+\Lambda^2}}
\end{equation}
is a stochastic increment which satisfies $\langle \delta x^2\rangle=2 D_b  \delta t$, $\langle \delta x\rangle=0$.

Similarly, the increment of $\phi$ may be expressed as
\begin{eqnarray}
\label{eq: 2.16}
\delta \phi&=&\frac{2\sqrt{\Lambda^2+1}}{\lambda_2(\Lambda^2-1)}\delta y
\nonumber \\
&=&\frac{1}{R}\frac{2(\Lambda^2+1)}{\Lambda^2-1}\delta y
\end{eqnarray}
where 
\begin{equation}
\label{eq: 2.17}
\delta y=\frac{\Lambda \delta u'_{1,1}+\delta u'_{1,2}}{\sqrt{\Lambda^2+1}}
\end{equation}
is a stochastic increment with the same statistics as $\delta x$, that
is $\langle\delta y^2\rangle=2 D_b  \delta t$.

The shape fluctuations are described by fluctuations of just two variables,
$\phi$ and $\Lambda$, which evolve diffusively. Their coefficients 
of diffusion and drift are independent of $\phi$:
\begin{eqnarray}
\label{eq: 2.18}
D_\phi (\Lambda)&=&\frac{4(\Lambda^2+1)^2}{(\Lambda^2-1)^2} D_b  
\nonumber \\
D_\Lambda(\Lambda)&=&(\Lambda^2+1)^2  D_b 
\nonumber\\
v_{\Lambda}(\Lambda)&=&\frac{2\Lambda(\Lambda^2+1)^2}{\Lambda^2-1} D_b 
\nonumber \\
v_\phi &=&0
\ .
\end{eqnarray}
An alternative coordinate is $z=2\Lambda/(\Lambda^2+1)$. Expressing 
the stochastic increments in the $(z,\phi)$ coordinate system, we obtain 
the following expressions for the diffusion and drift coefficients:
\begin{eqnarray}
\label{eq: 2.20}
D_z&=&\left(\frac{{\rm d}z}{{\rm d}\Lambda}\right)^2 D_\Lambda=4\left(\frac{\Lambda^2-1}{\Lambda^2+1}\right)^2 D_b 
\nonumber \\
v_z&=&\frac{{\rm d}z}{{\rm d}\Lambda}v_\Lambda +\frac{{\rm d}^2 z}{{\rm d}\Lambda^2} D_\Lambda
=\frac{-16\Lambda}{\Lambda^2+1} D_b 
\ .
\end{eqnarray}
Again following the approach in section \ref{sec: 3}, we express the diffusion and drift coefficients in 
terms of $z$ by introducing a variable $\Theta$ defined by 
$\Lambda=\sin\Theta/\cos\Theta$, so that $z=\sin(2\Theta)$.
Now express the coefficients in terms of $z$, and obtain
\begin{eqnarray}
\label{eq: 2.23}
D_z(z)&=&4\left(\frac{1-\Lambda^2}{1+\Lambda^2}\right)^2 D_b =\frac{4(\sin^2\Theta-\cos^2\Theta)^2}{(\sin^2\Theta+\cos^2\Theta)^2} D_b =4\cos^2(2\Theta) D_b =4(1-z^2) D_b 
\nonumber \\
D_\phi(z)&=&4\left(\frac{\Lambda^2+1}{\Lambda^2-1}\right)^2 D_b =\frac{16 D_b ^2}{D_z(z)}=\frac{4 D_b }{1-z^2}
\nonumber \\
v_z(z)&=&\frac{-16\Lambda}{\Lambda^2+1} D_b =-8z D_b 
\ .
\end{eqnarray}
These are the same as for simple diffusion on a sphere, as discussed in section 
\ref{sec: 1}, with the diffusion coefficient $D_b $ replaced by $4 D_b $. These coefficients 
are also very closely related to those obtained in the random strain model: the coefficients 
for that model all differ by a factor of $z^2$.  

\section{Mixed model}
\label{sec: 4}

Consider the distribution of triangle shapes for a mixed model, in which the displacement 
of the three particles in a short time interval $\delta t$ is a combination of 
a random strain and an uncorrelated random displacement, which is proportional to the 
radius of gyration $R$. This leads to the following map, using the notation of (\ref{eq: 0.1}):
\begin{equation}
\label{eq: 4.1}
\mbox{\boldmath$x$}'_i=
\mbox{\boldmath$x$}_i + \delta \mbox{\boldmath$x$}_i =
({\bf I}+\delta {\bf A})\mbox{\boldmath$x$}_i
+R\delta \mbox{\boldmath$r$}_i
\end{equation}
describing the evolution of the shape of the triangle, from $\mathbf{x}$ to
$\mathbf{x'}$ in a time $\delta t$. The statistics of the elements of
$\delta {\bf A}$ are given by (\ref{eq: 3.6}), and
$\delta \mbox{\boldmath$r$}_i$ has a Gaussian distribution, characterized
by $\langle \delta \mbox{\boldmath$r$}_i\rangle=0$ and 
$\langle \delta \mbox{\boldmath$r$}_i\cdot \delta\mbox{\boldmath$r$}_j\rangle=4D_{\rm b}\delta_{ij}\delta t$, 
where $D_{\rm b}$ is the diffusion coefficient associated with Brownian motion.
The scaling of the Brownian component in (\ref{eq: 4.1}) implies that the 
increments of $\Lambda$ and $\phi$ due to the Brownian motion are given 
by equations (\ref{eq: 2.14}) and (\ref{eq: 2.16}), alternatively by stochastic
incremenets $\delta z$, $\delta \phi$ which can be expressed using 
the drift and diffusion coefficients (\ref{eq: 2.23}). 
Similarly, there are independent contributions to $\delta z$  and $\delta \phi$ from the   
shearing motion. These are described by stochastic increments with drift and diffusion
coefficients given by equations (\ref{eq: 3.19}), (\ref{eq: 3.20}) and (\ref{eq: 3.21}).

The stochastic terms  $\delta \bf{A}$  and 
$\delta \mbox{\boldmath$r$}_i$ in (\ref{eq: 4.1}) both act multiplicatively, either 
on the variables $\mbox{\boldmath$x$}_i$, or upon the radius of gyration $R$, which is a function
of these variables. This could potentially result in
an ambiguity in defining the model \cite{vKa81}, depending upon whether we use 
the It\=o or Stratonovich definition of the stochastic increment. We notice in this respect 
the ambiguity in the evolution defined by the Brownian term
only affects the radial variable $R$, and not the 
shape variable. 
As pointed out in Section \ref{sec: 2}, the same is true for the shear term.

The fluctuations of shape are therefore described by a Fokker-Planck equation
for the joint probability density on the Kendal sphere, $P(z,\phi,t)$
\begin{equation}
\label{eq: 4.2}
\frac{\partial P}{\partial t}=-\frac{\partial}{\partial z}\left[v_z(z)P\right]+
\frac{\partial^2}{\partial z^2}\left[D_z(z)P\right]+\frac{\partial^2}{\partial \phi^2}
\left[D_\phi(z)P\right]
\end{equation}
with 
\begin{eqnarray}
v_z(z)&=&-8D_{\rm b}z-8D_{\rm s}z^3
\nonumber \\
D_z(z)&=&4(1-z^2)D_{\rm b}+4z^2(1-z^2)D_{\rm s}
\nonumber \\
D_\phi(z)&=&\frac{4D_{\rm b}}{1-z^2}+\frac{4D_{\rm s}z^2}{1-z^2}
\ .
\end{eqnarray}
The steady state solution only depends upon the ratio of diffusion
coefficients
\begin{equation}
\label{eq: 4.3}
\epsilon=\frac{D_{\rm b}}{D_{\rm s}}
\ .
\end{equation}
This has a steady state solution where $\phi$ is uniform on the circle and
where the marginal distribution for $z$ satisfies
\begin{equation}
\label{eq: 4.4}
2z(\epsilon+z^2)P+\frac{\partial}{\partial z}\left[(\epsilon+z^2)(1-z^2)P\right]=0
\end{equation}
so that 
\begin{equation}
\label{eq: 4.5}
2zP+(\epsilon+z^2)\frac{\partial P}{\partial z}=0
\ .
\end{equation}
This has a solution
\begin{equation}
\label{eq: 4.6}
P(z)=\frac{\sqrt{\epsilon}}{2{\rm tan}^{-1}(1/\sqrt{\epsilon})}\frac{1}{\epsilon+z^2}
\end{equation}
where the multiplier normalises $P(z)$ on $[-1,1]$. 

To check the prediction of (\ref{eq: 4.6}), we have studied numerically
the mixed model defined by (\ref{eq: 4.1}). One of the crucial properties of 
the random strain is that it conserves exactly the area of the triangle:
${\rm det}( \bf{M} ) = 1$, where $\bf{M}$ is introduced in 
(\ref{eq: 3.2},\ref{eq: 3.3}). 
We stress that the constraint ${\rm det}( \bf{M} ) = 1$ is satisfied only 
when the evolution equation (\ref{eq: 4.1}) is treated using the Stratonovich 
formulation~\cite{vKa81}. One readily notices that the expression for the 
evolution operator,
induced by the strain, and correct up to order $\delta t^1$, is simply:
\begin{equation}
\label{eq:exp}
\mbox{\boldmath$x'$}_i=
({\bf I}+\delta {\bf A} + \delta {\bf A}^2/2 )\mbox{\boldmath$x$}_i
\end{equation}
which is the expansion of the exponential of the traceless matrix 
$\delta {\bf A}$ up to second order. To impose strictly the constraint of 
area conservation, we replaced the operator in (\ref{eq:exp}) by 
$\exp( \delta {\bf A} )$, whose determinant is exactly equal to $1$.
and has the correct expansion, up to terms of order $\delta t^{3/2}$.
Using the fact that:
$\delta {\bf A}^2=( \delta a^2 + \delta b^2) \bf{I}$, one readily
finds that 
\begin{equation}
\exp( \delta {\bf A} ) = \cosh( \sqrt{(\delta a^2 + \delta b^2)} ) \bf{I}
+ \frac{\sinh( \sqrt{ (\delta a^2 + \delta b^2) } )}{\sqrt{ (\delta a^2 + \delta b^2 ) } } \delta {\bf A}
\ .
\label{eq: 4.7}
\end{equation}
Also, the Brownian motion term acts in (\ref{eq: 0.1})
multiplicatively (because of $R$). As we are interested here only in 
the shape evolution, and not on the evolution of the size, $R$, which 
requires some care, we have projected the term 
$R \delta \mbox{\boldmath$r$}_i  $ term in the 
direction transverse to the radial direction, which does not affect the 
shape dynamics, but simplifies the numerical integration. 

The data shown in Figure 1 was obtained by iterating the mapping
given by (\ref{eq: 4.1}) for $3 \times 10^4$ time steps, over 
$4 \times 10^5$ different realisations.
Figure 1 shows our
numerical results, which agree very well with the theoretical prediction.

\begin{figure*}
\label{fig:1}
\begin{center}
\epsfig{file=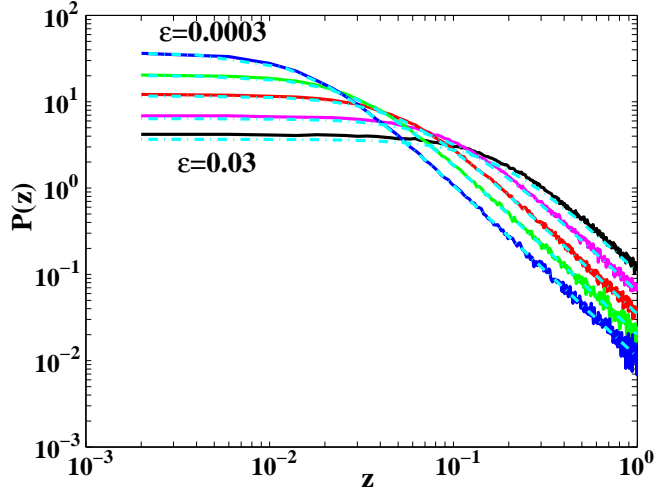,width=0.6\linewidth,clip=}
\caption{The PDF of $z$ for values of 
$\epsilon \equiv D_{\rm b}/D_{\rm s}$,
$\epsilon = 3\times10^{-4}$, $10^{-3}$, $3\times 10^{-3}$, $10^{-2}$ and $3\times10^{-2}$,
shown as a full line. The fit by equation (\ref{eq: 4.6})
is shown as a dashed line.  }
\end{center}
\end{figure*}

\section{Alternative formulation using Casimir operators}
\label{sec: 5}

The derivation of the stationary shape distribution in sections
\ref{sec: 2}-\ref{sec: 4} proceeded by setting up Langevin equations
for the evolution of the variables $z$ and $\phi$ (which label positions on the 
Kendall sphere of triangle shapes). This Langevin process was used
to obtain a one-dimensional Fokker-Planck equation for $z$, after
first concluding that the probability density is uniform in $\phi$. 
An alternative approach is to derive a Fokker-Planck equation for the 
distribution of all of the variables $z$, $\phi$, $\theta$ and $R$ which parametrise
the relative disposition of three points in the plane, and to solve the resulting
four-dimensional Fokker-Planck equation by separation of variables.
This alternative method 
makes plain the invariance properties under the group of
rotation, $SO(2)$ and the group of dilation, and in the case of the 
random strain model, under
the group of unitary matrices, $SL(2)$ \cite{Shr+98}, at the cost of
complicated algebraic manipulations,
which we do not present in full detail here. 
These invariances are expressed via the introduction of 
the Casimir operators, $\hat{\bf L}^2$, $\hat{\bf G}^2$ and $\hat \Lambda$
corresponding respectively to $SO(2)$, $SL(2)$ and the group of dilation
\cite{Shr+98}.
Consistent with the convention introduced in \cite{Shr+98}, we use
for the purposes of this section the 
coordinate $\zeta={\rm det}({\bf w})$ to characterise the size of the 
triangle, rather than the radius of gyration $R$. In fact, triangles
can be parametrized either by 
the set of coordinates 
$(z,\phi,\chi,\zeta)$ coordinates, as in \cite{Shr+98,Cas+01}, or by the 
set of coordinates $(z,\phi,\chi,R)$ used so far. Note that our notation 
differ from those of \cite{Shr+98} and
\cite{Cas+01}, where the coordinates $(z,\phi,\chi,\zeta)$
were denoted $(w,\chi,\theta,\zeta)$ (we changed the notation 
because it is conventional to use $(z,\phi)$ as the polar projection and 
azimuthal angle on a sphere).

\subsection{Fokker-Planck equation for random strain model}
\label{sec: 5.1}

Here we consider how to describe the random strain process using a Fokker-Planck
equation. We parametrize a triangle by the two vectors, $\mbox{\boldmath$u$}_a$, $a = 1,2$,
introduced in equation (\ref{eq: 1.1}). These vectors are specified 
by four coordinates $u_{a,i}$, and we start by describing the random strain
process by a Fokker-Planck equation expressed in terms of these four variables. 
Under the action of the random
strain process, these vectors evolve according to 
\begin{equation}
\label{eq: 5.1}
\dot{\mbox{\boldmath$u$}_a} = {\bf a}(t) \mbox{\boldmath$u$}_a 
\end{equation}
where ${\bf a}(t)$ is a random $2\times 2$ matrix.
The antisymmetric part of ${\bf a}(t)$ corresponds to an overall
rotation. Since we are interested only in the shape, we ignore the 
overall rotation and dilation, and consider only traceless 
symmetric matrices.

By integrating equation (\ref{eq: 5.1}) with respect to time and iterating 
the resulting 
equation, in the limit of small values of
the time $\delta t$ one obtains:
\begin{equation}
\label{eq: 5.2}
u_{a,i}(\delta t) = u_{a,i}(0) + \int_0^{\delta t} {\rm d}t'\ a_{i,j}(t') u_{a,j}(0)
+ \int_0^{\delta t} {\rm d}t' \int_0^{t'} {\rm d}t'' a_{ik}(t') a_{k,j}(t'') u_{a,j}(0) + 
{\cal O}(\delta t^{3/2})
\label{eq:int_2}
\end{equation}
(here and in the remainder of this section summation is implied when indices are repeated). 
In equation (\ref{eq:int_2}), the term written as a single integral has 
zero mean, and a variance 
growing like $\delta t$. It is therefore a diffusion term. The term 
written with a double 
integral has a mean of order $\delta t$; it corresponds to the drift term. 
Higher order
terms are formally smaller order terms in $\delta t$. To make the connection with the formulation used
in section \ref{sec: 3}, the stochastic increment in equations (\ref{eq: 3.4}) 
and (\ref{eq: 3.5}) may be written
\begin{equation}
\label{eq: 5.2.5}
\delta {\bf A}=\int_t^{t+\delta t}{\rm d}t'\ {\bf a}(t')
\ .
\end{equation}

We assume that the coefficients of the matrix ${\bf a}(t)$ are 
$\delta$-correlated in time. 
Using the constraint that ${\bf a}$ is symmetric, traceless, and has 
rotationally invariant statistics, we find
\begin{equation}
\label{eq: 5.3}
\langle a_{ij}(t) a_{kl}(0) \rangle = \langle a^2 \rangle \times \delta(t) 
\times 
( \delta_{ik} \delta_{jl} + \delta_{il} \delta_{jk} - \delta_{ij} \delta_{kl} )
\label{eq:variance_a}
\end{equation}
where $\langle a^2 \rangle$ is the only parameter that characterizes the 
amplitude of the noise. This relation is equivalent to equations (\ref{eq: 3.5})
and (\ref{eq: 3.6}), with $\langle a^2\rangle=2D_s$.
The use to the relation (\ref{eq:int_2}) for advancing the variables
$u_{a,i}$ amounts to using the physically correct Stratonovich procedure 
to integrate the
stochastic model (\ref{eq: 0.1}).

From equation (\ref{eq:int_2}) and (\ref{eq:variance_a}) one can 
easily determine the Fokker-Planck equation for the random-strain process:
\begin{eqnarray}
\label{eq: 5.4}
\partial_t P &=& - \frac{\partial}{\partial u_{a,i}} (v_{a,i} P) + \frac{ \langle a^2 \rangle}{2}
\frac{\partial^2}{\partial u_{a,i} \partial u_{b,k}} 
\Bigl( [ \delta_{ik} \mbox{\boldmath$u$}_a(0) \cdot \mbox{\boldmath$u$}_b(0)
+ u_{a,k}(0) u_{b,i}(0) - u_{a,i}(0) u_{b,k}(0) ] P \Bigr)
\nonumber \\
&\equiv &  - \frac{\partial}{\partial u_{a,i}} (v_{a,i} P) + \frac{ \langle a^2 \rangle}{2}\hat L_d P
\label{eq:FP_incompl}
\end{eqnarray}
where $v_{a,i}$ is the drift term velocity, obtained by taking the average of (\ref{eq: 5.2}), 
and $L_d$ is a diffusion operator containing second derivatives. For the drift term we find
\begin{equation}
\label{eq: 5.5}
v_{a,i} = {\langle a^2 \rangle} u_{a,i}
\ .
\end{equation}
The term representing the diffusion term in the Fokker-Planck operator has a more complex
structure. We find it convenient to write $L_d$ as the sum of three operators,
\begin{eqnarray}
\label{eq: 5.7}
\hat L_d^1 \cdot P & \equiv & \frac{\partial^2}{\partial u_{a,i} \partial u_{b,i}} [ \mbox{\boldmath$u$}_a(0) \cdot \mbox{\boldmath$u$}_b(0) P ] 
 \label{eq:diff_op1} \\
\hat L_d^2 \cdot P & \equiv & \frac{\partial^2}{\partial u_{a,i} \partial u_{b,j}} [ u_{a,j}(0) \cdot u_{b,i}(0) P ]  
\label{eq:diff_op2} \\
\hat L_d^3 \cdot P & \equiv & \frac{\partial^2}{\partial u_{a,i} \partial u_{b,j}} [ u_{a,i}(0) \cdot u_{b,j}(0) P ]
\ . 
\label{eq:diff_op3}
\end{eqnarray}

\subsection{Representation using Casimir operators}
\label{sec: 5.2}

Now consider how the Fokker-Planck equation (\ref{eq: 5.4}) may be transformed 
from Cartesian coordinates to the more convenient set $(z,\phi,\chi,\zeta)$ which
separate the actions of shape, rotation, and scale transformations. As a 
consequence, we shall see how the operator may also be decomposed into Casimir
operators for these fundamental group operations. 

The that drift term in the Fokker-Planck equation, $\partial_a^i (v_a^i P) $, makes a 
contribution which takes a very simple form, involving only the coordinate $\zeta$:
\begin{equation}
\label{eq: 5.6}
\frac{\partial}{\partial u_{a,i}} (v_{a,i} P) = 2 \langle a^2 \rangle 
\Bigl( 2 \zeta \frac{ \partial P}{\partial \zeta} + 4 P \Bigr)
\label{eq:drift-term}
\end{equation}
Each of the three  operators in equation (\ref{eq: 5.7}) can be written in terms of 
the Casimir operators of $SO(2)$ ($\hat {\bf L}^2 $), of $SL(2)$
($\hat {\bf G}^2$) and of the dilation operator, $\hat\Lambda$.

The angular momentum operators are, up to a factor $\pm{\rm  i}$,
\begin{equation}
\label{eq: 5.8}
\hat L^{ij}.P = ( u_{a,i} \frac {\partial }{\partial u_{a,j}} - u_{a,j}\frac{\partial}{\partial u_{a,i}} ) P 
\end{equation}
and in this two-dimensional problem the corresponding Casimir operator is 
\begin{equation}
\label{eq: 5.9}
\hat {\bf L}^2 = \frac{1}{2}\hat  L^{ij}\hat  L^{ji}
\ .
\end{equation}
A straightforward calculation leads to:
\begin{equation}
\label{eq: 5.10}
\hat {\bf L}^2 P = \frac{\partial^2}{\partial u_{a,i} \partial u_{b,j}} ( u_{a,j} u_{b,i} P) 
- \frac{\partial^2}{\partial u_{a,i} \partial u_{b,i}} ( \mbox{\boldmath$u$}_a \cdot \mbox{\boldmath$u$}_b P ) 
- ( d - 1) ( d^2 + \hat \Lambda ) P
\label{eq:expr-L2}
\end{equation}
where $d = 2$ is the space dimension, and where
\begin{equation}
\label{eq: 5.11}
\hat \Lambda = u_{a,i} \frac{\partial}{\partial u_{a,i}}
\ .
\end{equation}

The operators that generate the representations of $SL(2)$ are:
\begin{equation}
\label{eq: 5.12}
\hat G^{ab} = u_{a,i} \frac{\partial}{\partial u_{b,i}} - \frac{\delta_{ab}}{2} \hat \Lambda
\ .
\label{eq:def_G}
\end{equation}
The operator $G^{ab}$ and $\Lambda$ commute with each other.

The operator ${\bf G}^2 \equiv G^{ab} G^{ba}$ can be easily expressed in 
the form:
\begin{equation}
\label{eq: 5.13}
\hat {\bf G}^2 = \frac{\partial^2}{\partial u_{b,i} \partial u_{a,j}} ( u_{a,i} u_{b,j} \cdot ) 
- \frac{1}{d} (  \hat\Lambda + d^2 ) (\hat \Lambda + 2 d^2)
\ .
\label{eq:G2}
\end{equation}
Last, we express $\hat \Lambda^2$:
\begin{equation}
\label{eq: 5.14}
\hat \Lambda^2 = \frac{\partial^2}{\partial u_{a,i} \partial u_{b,j}} ( u_{a,i} u_{b,j} \cdot ) 
- (2 d^2 + 1) (d^2 + \hat\Lambda) + d^4 
\ .
\label{eq:Lambda2}
\end{equation}
Thus, the 3 operators $\hat L_d^i$ are:
\begin{eqnarray}
\label{eq: 5.15}
\hat L_d^1  & = & - \hat{\bf L}^2 + \hat{\bf G}^2 
+ \frac{1}{d} (\hat\Lambda + d^2 ) (\hat\Lambda + 2 d^2 ) - (d - 1) (\hat\Lambda + d^2 ) 
\label{eq:exp_L1} \\
L_d^2 & = & \hat{\bf G^2} 
+ \frac{1}{d} (\hat\Lambda + d^2 ) (\hat\Lambda + 2 d^2 )
\label{eq:exp_L2} \\
L_d^3 & = & \hat\Lambda^2 + (2 d^2 + 1) (\hat\Lambda + d^2) - d^4 
\label{eq:exp_L3}
\ .
\end{eqnarray}
With these expressions, one can simply obtain the expression for $\hat L_d$:
\begin{equation}
\label{eq: 5.16}
\hat L_d \cdot U = ( - \hat{\bf L}^2 + 2 \hat{\bf G}^2 + 2 \hat\Lambda + 8 ) U
\ .
\label{eq:full_op}
\end{equation}
The terms $\hat{\bf L}^2$, $\hat{\bf G}^2$ and $\hat\Lambda^2$ all have a simple expression
in terms of the variables $(z,\phi,\chi,\zeta)$ \cite{Shr+98}.
Specifically,
\begin{eqnarray}
\label{eq: 5.17}
\hat {\bf L}^2 & = & \frac{\partial^2}{\partial \chi^2} 
\label{eq:L2_coor} \\
\hat {\bf G}^2 & = & 2 z^2 \frac{\partial}{\partial z} \left[ ( 1 - z^2 ) \frac{\partial}{\partial z}\right]   + \frac{z^2}{2 (1 - z^2) } \Bigl( 
\frac{\partial^2}{\partial \chi^2} + 
\frac{\partial^2}{\partial \phi^2} - 2 z \frac{\partial^2}{\partial \chi \partial \phi} \Bigr) 
\label{eq:G2_coor} \\
\hat \Lambda^2 & = & 4 \left( 
\zeta^2 \frac{\partial^2}{\partial \zeta^2} + 
\zeta \frac{\partial}{\partial \zeta} \right)
\ .
\label{eq:Lambda2_coor}
\end{eqnarray}
Thus, the operator $\hat L_d$ reads:
\begin{eqnarray}
\label{eq: 5.18}
\hat L_d \cdot U & = &  4 z^2 \frac{\partial}{\partial z}\left[ ( 1 - z^2 ) \frac{\partial U }{\partial z}\right ]
+  \frac{z^2}{ (1 - z^2) } \Bigl( \frac{\partial^2}{\partial \phi^2} - 2 z \frac{\partial^2}{\partial \chi \partial \phi} \Bigr) U  
\nonumber \\
& + & \frac{1}{1 - z^2 } \frac{\partial^2 U}{\partial \phi^2}
+ 4 \zeta \frac{\partial U}{\partial \zeta}  + 8 U 
\ .
\label{eq:Ld_coor} 
\end{eqnarray}

\subsection{Relation to formulation in section \ref{sec: 3}}
\label{sec: 5.3}

Equation (\ref{eq: 5.18}) enables us to express the Fokker-Planck operator
in terms of the coordinates $(z,\phi,\chi,\zeta)$. Let us consider how to 
obtain the results developed in section \ref{sec: 3} using this representation of the
Fokker-Planck operator. Because we intend to look at solutions which are independent of the rotation
angle $\chi$, the
diffusion part of the operator is: 
\begin{equation}
\label{eq: 5.19}
L_d \cdot U =  4 z^2 \frac{\partial}{\partial z}\left[ ( 1 - z^2 ) \frac{\partial U }{\partial z} \right]
+  \frac{z^2}{ (1 - z^2) }  
\frac{\partial^2 U}{\partial \phi^2} 
+ 4 \zeta \frac{\partial U}{\partial \zeta}  + 8 U
\ . 
\label{eq:Ld_simpl} 
\end{equation}
Thus, the full Fokker-Planck equation, expressed as in equation (\ref{eq:FP_incompl}),
is simply
\begin{equation}
\label{eq: 5.20}
\partial_t P = \langle a^2 \rangle
\Bigl( 2 z^2 \frac{\partial}{\partial z} \left[ ( 1 - z^2 ) \frac{\partial P }{\partial z}\right ]  
+  \frac{z^2}{ 2 (1 - z^2) }  \frac{\partial^2 P}{\partial \phi^2}  \Bigr)
\ .
\label{eq:FP-RSM}
\end{equation}
As expected, this is independent of $\zeta$. However, its structure appears
hard to reconcile with the results for the random strain model in section \ref{sec: 3}.
In order to make the connections clear, we first remark that this is an equation for a 
probability density $P$ expressed as a function of the four original variables, 
$u_{a,i}$, $a=1,2$, $i=1,2$. The transformation to the new variables
$\zeta,\chi,z,\phi$ involves the Jacobian
\begin{equation}
\label{eq: 5.21}
\frac{D( u_{1,1}, u_{1,2} , u_{2,1}, u_{2,2} ) }{D(z,\phi,\chi,\zeta) }
= \frac{\zeta}{z^2}
\ .
\label{eq:jacobian_w_zeta}
\end{equation}
The total probability is obtained by integrating $P$ over all
the $u_{a,i}$. After transforming to the $(z,\phi,\chi,\zeta)$ variables
it is more convenient to deal with a probability density $P'(z,\phi,\chi,\zeta)$ which
 is a conserved quantity in these variables.  In the $(z, \phi, \chi, \zeta)$ variables, 
the normalization condition reads:
\begin{eqnarray}
\label{eq: 5.22}
&&\int {\rm d}u_{1,1}\,{\rm d}u_{1,2}\,{\rm  d}u_{2,1}\,{\rm d}u_{2,2}\  P(u_{1,1}, u_{1,2}, u_{2,1},u_{2,2} )
\nonumber \\
&=& \int {\rm d}\zeta {\rm d}\chi\,{\rm d}z \,{\rm d}\phi\  P(\zeta, \chi,z,\phi) 
\frac{\zeta}{z^2}=\int {\rm d}\zeta\, {\rm d}\chi\, {\rm d}z\, {\rm d}\phi\  P'(\zeta, \chi,z,\phi) 
\ .
\label{eq:use_Jacob}
\end{eqnarray}
If we rewrite equation (\ref{eq:FP-RSM}) in terms of $P'$ we obtain
\begin{equation}
\label{eq: 5.23}
\partial_t P' = \langle a^2 \rangle
\Bigl( 2 \frac{\partial}{\partial z} \left[ ( 1 - z^2 ) \frac{\partial z^2 P' }{\partial z}\right]
+  \frac{z^2}{ 2 (1 - z^2) }  \frac{\partial^2 P'}{\partial \phi^2}  \Bigr)
\ .
\label{eq:FP-RSM-v2}
\end{equation}
Using some elementary manipulation of the part in $z$ of the 
evolution operator, one can rewrite:
\begin{equation}
\label{eq: 5.24}
\frac{\partial}{\partial z} \left[ ( 1 - z^2 ) \frac{\partial z^2 P' }{\partial z} \right]
= \frac{\partial^2}{\partial z^2} \left[ z^2 ( 1-z^2) P'\right] - 
\frac{\partial}{\partial z} (- 2 z^3 P )
\label{eq:identity_1}
\end{equation}
which agrees with the conclusion that the change in shape can 
be described by a diffusion $ z^2 (1 - z^2)$ and a drift $ -2 z^3$.

\subsection{Fokker-Planck equation for Brownian process}
\label{sec: 5.4}

Next consider the Fokker-Planck equation for the relative Brownian
diffusion of three points, and its relation to the derivation
of Kendall's result on diffusion which as discussed in section 
\ref{sec: 2}. After a lengthy calculation, this is found to 
take the following form
\begin{eqnarray}
\label{eq: 5.25}
\partial_t P = \frac{D_{\rm b}}{z \zeta} & \Bigl[ & 
2 \zeta^2 \frac{ \partial^2 P}{\partial \zeta^2} 
+ 2 z^2 \partial_z [ (1 - z^2) \partial_z P ] 
\nonumber \\
& + & \frac{z^2}{2 (1 - z^2) } 
\left( \frac{\partial^2 P}{\partial \phi^2} 
+ \frac{\partial^2 P}{\partial \chi^2} 
- 2 z \frac{\partial^2 P}{\partial \chi \partial \phi} \right) \nonumber \\
& + & 4 (1 - z^2) z \zeta  \frac{\partial^2 P}{\partial z \partial \zeta } ~~~\Bigr]
\end{eqnarray}
where $D_{\rm b}$ is the amplitude of the additive noise term:
\begin{equation}
\label{eq: 5.26}
\partial_t u_{a,i} = v_{a,i} ~~~ {\rm with } ~~~ \langle v_{a,i}(t) v_{b,j}(0) \rangle = 2 D_{\rm b} 
\delta(t) \delta_{ab} \delta_{ij}
\ .
\label{eq:add_noise}
\end{equation}
Because we chose to consider a process that does not change the radius of the solution. 
This suggests that the evolution will be simple in the variables $(z, \phi, \chi, R)$. Indeed,
the evolution operator in these variables is simply:
\begin{equation}
\label{eq: 5.27}
\partial_t P = D_{\rm b} \times \Bigl(
\frac{\partial}{\partial z} \left[ (1 - z^2) \frac{\partial}{\partial z} P \right] 
+ \frac{1}{1-z^2} ( \partial_{\phi^2} + \partial_{\chi^2} - 2 z \partial_{\phi} \partial_{\chi} )P  \Bigr) 
\label{eq:lap_rho}
\end{equation}
The Jacobian of the transformation to the variables $(z,\phi,\chi,R)$ is  proportional to $R^3$,
 i.e., it does not depend on the angular variables. The simple identity:
\begin{equation}
\label{eq: 5.28}
\frac{\partial}{\partial z} \left[ ( 1 - z^2 ) \frac{\partial P }{\partial z}\right]  
= \frac{\partial^2}{\partial z^2} \left[ ( 1-z^2) P\right]  - 
\frac{\partial}{\partial z} (- 2 z P )
\label{eq:identity_2}
\end{equation}
immediately leads to the conclusion that the evolution of $z$ results
from a diffusion  (coefficient $\propto ( 1 - z^2)$) and 
a drift (coefficient $\propto - 2 z$), as obtained in section \ref{sec: 2}.

\section{Summary}
\label{sec: 6}

We have analysed the simplest model which could describe the shape statistics 
for triplets of points advected in a two-dimensional random flow. The model combines 
Brownian processes which describe a random strain with a Brownian motion
of the triangle vertices. We characterised
the shape using Kendalll's spherical representation of the manifold of triangle shapes.
The steady-state distribution is uniform in the azimuthal coordinate $\phi$ 
but has a Lorentzian form in the polar altitude $z$, depending upon a single
dimensionless parameter. 

The model considered in this work can be viewed as a simplification 
of the models proposed in \cite{Cas+01,PSC+00}, which are taking into
account the hierarchy of time scales in a turbulent flow. Understanding
the shape distributions predicted by these models will require some 
generalization of the ideas developed in this work. Still, we expect
that the solution discussed here provides a starting point for future analysis
of the shape distribution of triangles, or more complicated sets of points 
transported by turbulent flows.

{}

\end{document}